# Reduction in Rebound of Concrete Piles Driven into Clays by Coating Pile Surface with Titanium Dioxide Nanoparticles

Nadya Amalia[1], Asifa Asri[1], Mamat Rokhmat[1], Sutisna[1], and Mikrajuddin Abdullah[1*]

**Abstract**

Using a model for concrete piles driven into clays, we compared penetration depths between uncoated piles and piles coated with titanium dioxide ($TiO_2$) nanoparticles. The behavior of surfaces coated with $TiO_2$ changes to superhydrophilic, enabling water molecules to penetrate inside the clay pores. The attraction suppresses or reduces the compression of water inside the clay pores. The absence of bulk pressure from water causes the pile not to bounce (backward movement after striking). Contrary to hydrophobic surfaces, which tend to repel water molecules, water is compressed into the clay pores generating a bulk pressure that induces a countering upward force (resulting in rebounding). Driving tests for two types of clay demonstrate the absence of bouncing from coated piles. An examination of the pile surfaces indicates the formation of bonds between water molecules and coated surface and the absence of such bonding for uncoated piles. This finding might accelerate the process of pile driving at any civil engineering construction site.

-----------------------------------------------------------------------

[1] Department of Physics, Bandung Institute of Technology, Jalan Ganesa 10 Bandung 40132, Indonesia, Tel. +62-22-2500834, Fax. +62-22-2506452, *E-mail: din@fi.itb.ac.id

**Introduction**

One interesting phenomenon commonly observed during pile driving is the backward movement of the pile after striking. This phenomenon is commonly known as high pile rebound (HPR) or bouncing [1,2,3]; the displacement can exceed 0.25 inches. HPR may affect drivability [4] and complicate the determination of the load bearing capacity of the pile [5]. This problem usually occurs if piles are driven into silts and clays [1,2,5]. There are several reviews on HPR at similar locations in North America [3,5]. Studies of piles driven into clays along the North Carolina coast indicated the frequent occurrence of HPR [3]. When excess pressure is applied to the pile to overcome HPR, the pile may experience circumferential cracking.

Pile driving may increase the water pressure in soil pores [6] because of the incompressible behavior of water and the hydrophobic behavior of the pile surface. Chandler reported the development of a gap or density depletion at the interface of water and hydrophobic surface [7]. Poynor et al. reported that the depletion region of 2 – 4 Å thickness might be created when water molecules are in contact with the hydrophobic surface [8]. From the relationship between the contact angle of a water droplet on a hydrophobic surface and the thickness of the depletion layer (Figure 1) [9], we find a thickness of between 1–8 Å for a contact angle in the range of 105–180°.

We can estimate the force during bouncing experienced by the pile based on the thickness of the depletion layer. Let us assume the diameter of the pile is $D = 0.2$ m and the bulk modulus of water is $B = 2\times10^9$ N/m$^2$. The formation of the depletion layer $\delta \approx$ 2–4 Å [8] implies a compressibility of water over the same distance. The volume of the pore containing water in the clay is in order of a cubic micrometer (the pore diameter $d \approx 1$ μm). We can estimate the repulsive force resulting from the presence of the depletion layer (compressibility of water) to be around $F \approx B (\Delta V/V)A$. With $A = \pi D^2/4$, $V \approx \pi d^3/6$, and $\Delta V \approx (\pi d^2/4)\delta$, we obtain $F \approx 6B\delta D^2/\pi d \approx$ 30,000 N. This value, however, might be overestimated as that the force resulting from the incompressibility of water is very large.

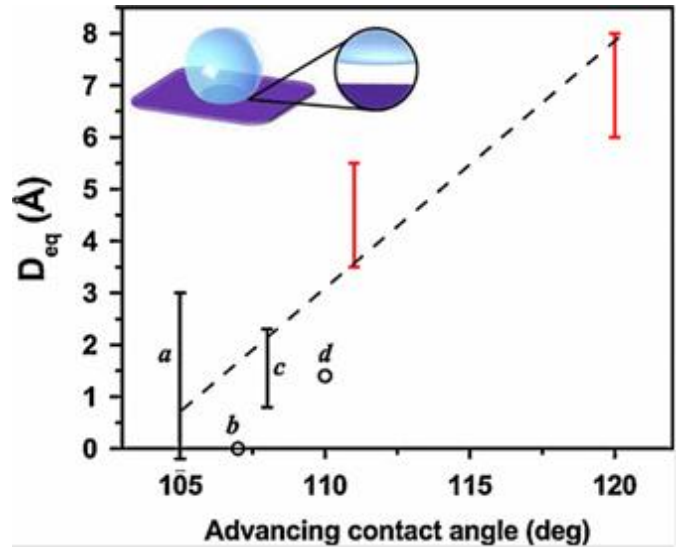

Figure 1. Relationship between the thickness of depletion layer on a hydrophobic surface with contact angle [9]. Reproduced with permission from American Physical Society.

Is it possible that the origin of the bouncing force is Archimedean buoyancy? This is given by $F = \rho g h A = \rho g h (\pi d^2/4)$, with $h$ is the depth of pile penetration. Using the mass density of water and a penetration depth of around 10 m, we estimate a force of 3,141 N. This value is very small compared with the force of incompressibility, indicating that the force responsible for pile bouncing is the force of incompressibility force for pore water.

If the pile surface exhibits hydrophilic behavior, it is easily wetted by water; therefore, the pile is attracted by the clay. In addition, hydrophilic materials also exhibit a small coefficient of friction [10]. Both effects, attraction between pile surface and clay and reduction in water pressure at the pile tip, reduce the bouncing force and stabilize the pile position after the pile has been driven in.

Theoretically, if the pressure of pore water at the pile tip (bottom) could be reduced during pile driving, HPR can be minimized or neglected. By considering the interaction of molecules on the pile surface with water, coating on the pile surface with materials that are capable of binding water molecules in the soil pores might be a solution to the HPR problem. Titanium dioxide ($TiO_2$) is a material with superhydrophilic properties under light illumination. This property persists for several hours after illumination has stopped. Superhydrophilic materials are able to bind water. Coating on the pile surface with superhydrophilic materials can be used to bind water molecules in the soil pore; therefore the repulsive force against the pile as a result of excess pore water pressure can be reduced.

**Methods**

We made specimen piles using a mixture of cement, sand, gravel, and water (with weight fractions of 1:1:1:0.5) in accordance with Indonesian National Standard (SNI) No. 03-2834-2000. A homogeneous mixture was cast into polyvinyl chloride (PVC) pipes of 30-cm length and different diameters: 1 inch, 1.25 inch, 1.5 inch, and 2 inch. The pile tip was made using a hard paper forming a cone of about 45° tip angle. An 8-mm-diameter iron rod was inserted along the pile axis.

After aging for one week, the pile molds (PVC and tip cover) were removed and the pile aged for a further week prior to coating. Some of the two-week-aged piles were then coated with $TiO_2$ particles (Bratachem, Indonesia); epoxy resin (araldite) was used as binding agent. We reported previously that the average size of the $TiO_2$ particles was 160 nm [11]. The mixture of epoxy resin and corresponding hardener (volume fraction of 1:1) was rubbed uniformly over the pile surface using a sponge and then TiO$_2$ particles poured over the surface. Some piles were coated using epoxy resin only (without TiO$_2$) to investigate the effect of epoxy on pile driving. The coated piles were then aged for a further two weeks giving a total aging period of four weeks. Pile driving tests were performed using a replica drop hammer (Figure 2). It was made of wood with total height of 180 cm, supported by three legs with angle of 30$^\circ$ with respect to vertical. The hammer was a concrete cylinder of diameter 10 cm, height 13 cm, and weight 3.2 kg.

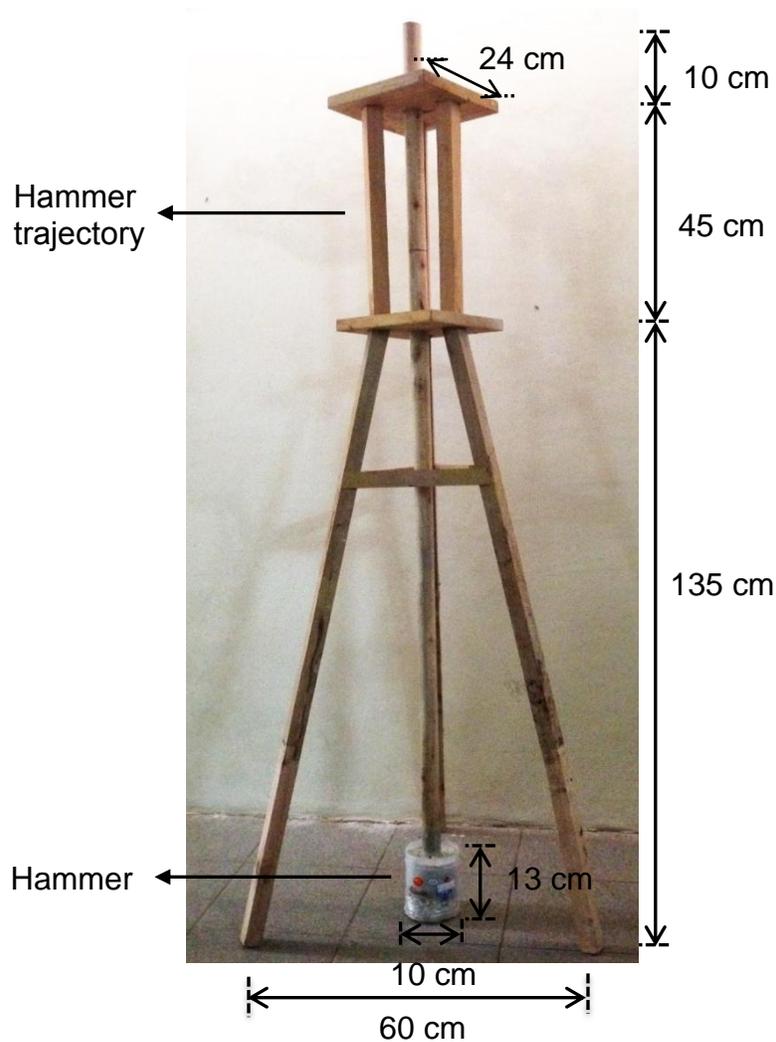

Figure 2. Schematic of the replica drop hammer.

We used clays obtained from two locations: Plered (Purwakarta area) and Sukabumi area, both are located in West Java Province, Indonesia. For comparison, we also used regular soil (non-clay) of different water content. Regular soil wet 1 was made by mixing regular soil with water at water/soil ratio of 1:2 vol%. Regular soil wet 2 was made by mixing 150 mL water with 2 kg of regular soil wet 1. Figure 3 shows a photograph of the soils used in pile driving tests.

We measured the water content in the clays, regular soil, regular soil wet 1, and wet 2 using a gravimetric meter (ASTM D2216). The mass of water is equal to the difference of wet soil mass and dry soil mass. The dry soil was dried in an oven at temperatures between 100–110 °C for 24 hours until a constant mass was achieved [12]. The soils were weighed and then dried again for 3 hours to ensure constant mass.

For the pile driving tests, the clays and regular soils were put inside a plastic container of 40-cm diameter and 40-cm height. Before testing, all samples were conditioned under the same properties. The tests were performed entirely on a single day to ensure no changes in water content.

The hydrophobicity and hydrophilicity of the pile surfaces were simply determined by measuring the contact angles of several water droplets on the pile surface. Using a video camera, measurements were made by recording the development of the contact angle. The final stable angles were assumed as true contact angles.

We also recorded the IR absorption using a Fourier transform infrared (FTIR) spectrometer (Bruker ALPHA) to analyze the bonding properties between the pile surface with clays and regular soils.

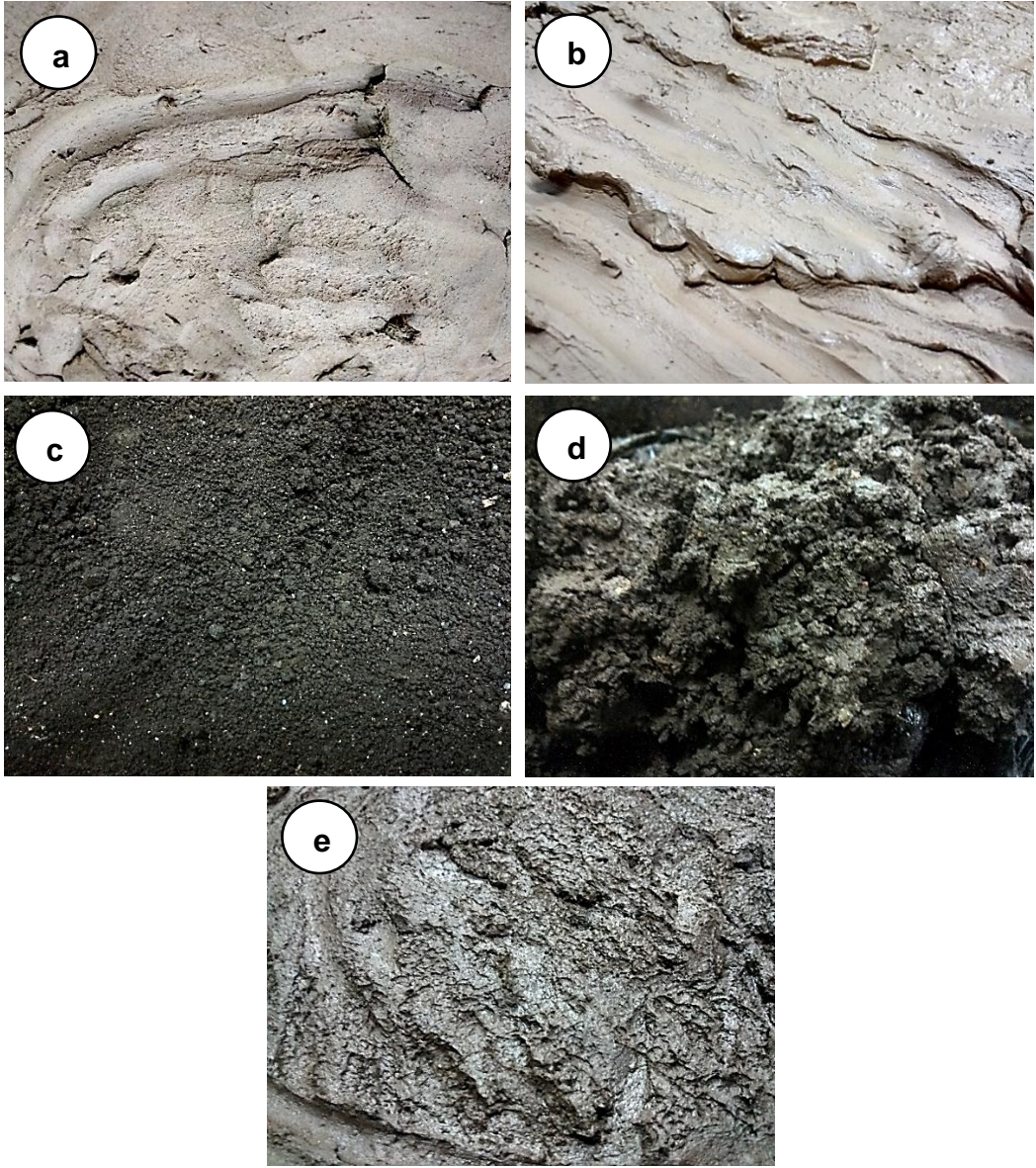

Figure 3. Photograph of soils used in pile driving tests: (a) Plered clay, (b) Sukabumi clay, (c) regular soil, (d) regular soil wet 1, and (e) regular soil wet 2.

**Results and Discussion**

Figure 4 are photos of examples of piles with and without a TiO$_2$ coating and with epoxy coating only. The piles shown are 1.25 inches in diameter; those of different diameters (1 inch, 1.5 inch, and 2 inch) are not shown.

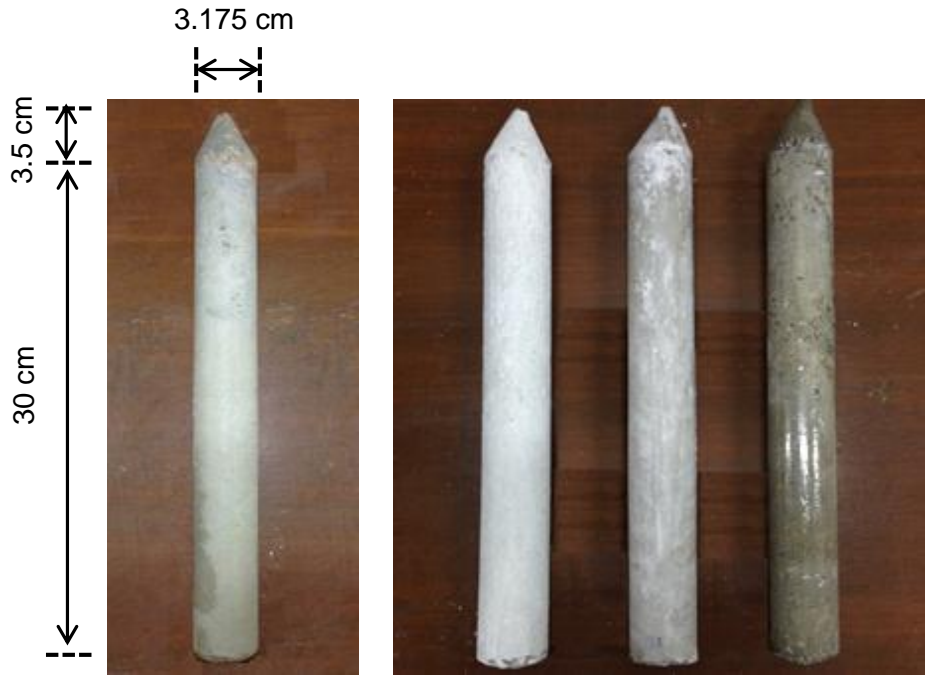

Figure 4. (left to right) Photographs of pile samples without coating (bare concrete), thick coating with $TiO_2$, with thin coating $TiO_2$, and coating with epoxy only. The diameter of each pile is 1.25 inch. The tips have a length of 3.5 cm so that the tip angle is around $45^o$.

As mentioned earlier, the hydrophobicity or hydrophilicity of the surface was measured by dropping water droplets on the pile surface. We recorded the development of the contact angles. Initially, contact angles were large but diminished with time until a stable value was reached after about 2 seconds. Measurements of several drops showed that roughly 10% of the piles had contact angles of larger than 90° and the rest had contact angles between $10^o$ - 90°. However, with a $TiO_2$ coating, the surface exhibited superhydrophilic behavior (Figure 5). Some recorded images are shown in the figure where initially the contact angle was very small and then vanished after 2 seconds.

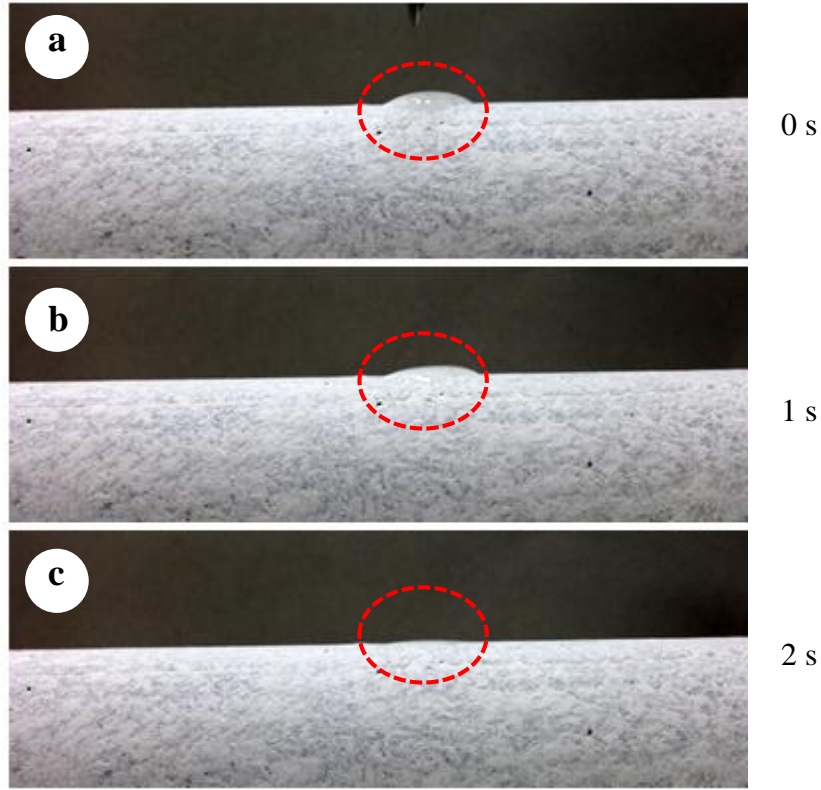

Figure 5. Development of contact angle of water droplet on a TiO$_2$-coated pile. The droplets are recorded at different times: (a) 0 sec, (b) 1 sec, and (c) 2 sec.

Table 1 shows the results of water content measurement for various soils. We measured four samples for each soil; the displayed results are averaged values. The dry weight water content was calculated using relation $\theta_d$ = (water mass/dry soil mass) × 100%; the wet weight water content was calculated similarly using $\theta_w$ = (water mass/wet soil mass) × 100% [12]. The water content of clays is larger than the water content of regular soils. Plered clay is slightly drier than Sukabumi clay. The regular soils, wet 1 and wet 2, have larger water content than the clays.

Table 1. Water content of clays and regular soils measured using a gravimetric meter

| Soils | Average $\theta_d$ (wt%) | Average $\theta_w$ (wt%) |
|---|---|---|
| Plered clay | 46.4 | 31.7 |
| Sukabumi clay | 49.6 | 33.1 |
| Regular soil | 42.7 | 29.9 |
| Regular soil added with less water (wet 1) | 56.6 | 36.1 |
| Regular soil added with more water (wet 2) | 66.2 | 39.9 |

Figure 6 shows a comparison of penetration depths in Plered clay of a pile without coating, with epoxy coating, and with $TiO_2$ coating. We used a 7-kg load positioned 60 cm above the top end of the pile, and applied a specified force three times. Following the above sequence of pile type, the penetrations depths were 21, 20, and 25 cm, respectively. These results indicate that epoxy resin has little effect on penetration depth. Only the $TiO_2$-coated pile was significantly different experiencing an additional penetration depth of about 20% compared with the uncoated pile.

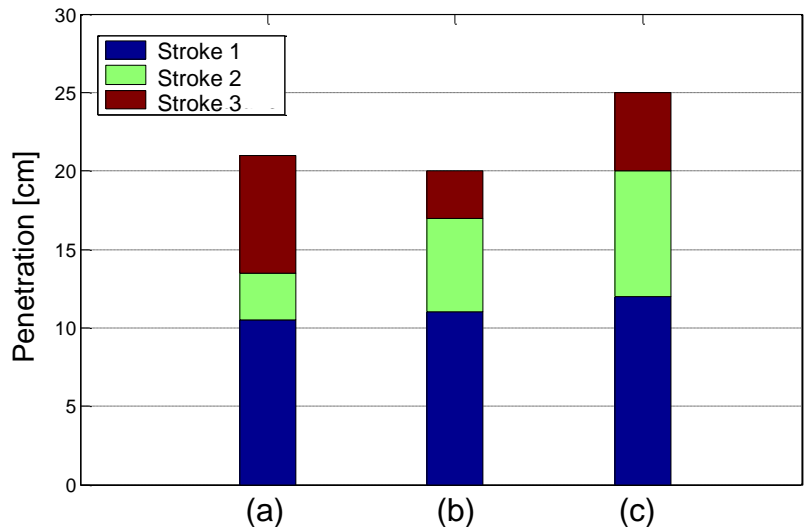

Figure 6. Penetration depths in Plered clay for: (a) uncoated, (b) epoxy-coated, and (c) $TiO_2$-coated piles. A 7-kg load positioned 60 cm above the pile top end was used to apply a constant force three times for each pile.

A comparison of penetration depths at each stroke for uncoated and $TiO_2$-coated piles in Sukabumi and Plered clays is shown in Figure 7. The tests were performed for piles of diameter 1.25 inch. The penetration depths are larger for $TiO_2$-coated piles than uncoated piles both in Sukabumi and Plered clays. With the difference being larger in Sukabumi clay, piles are more easily driven into this clay type. A 30-cm penetration was achieved in Plered clay after 14 strokes, whereas the same penetration could be achieved in Sukabumi clay after only 9 strokes. From Table 1, Sukabumi clay contains more water than Plered clay and hence we may conclude that a pile would penetrate easier and the difference in penetration depth between uncoated and $TiO_2$-coated becomes larger.

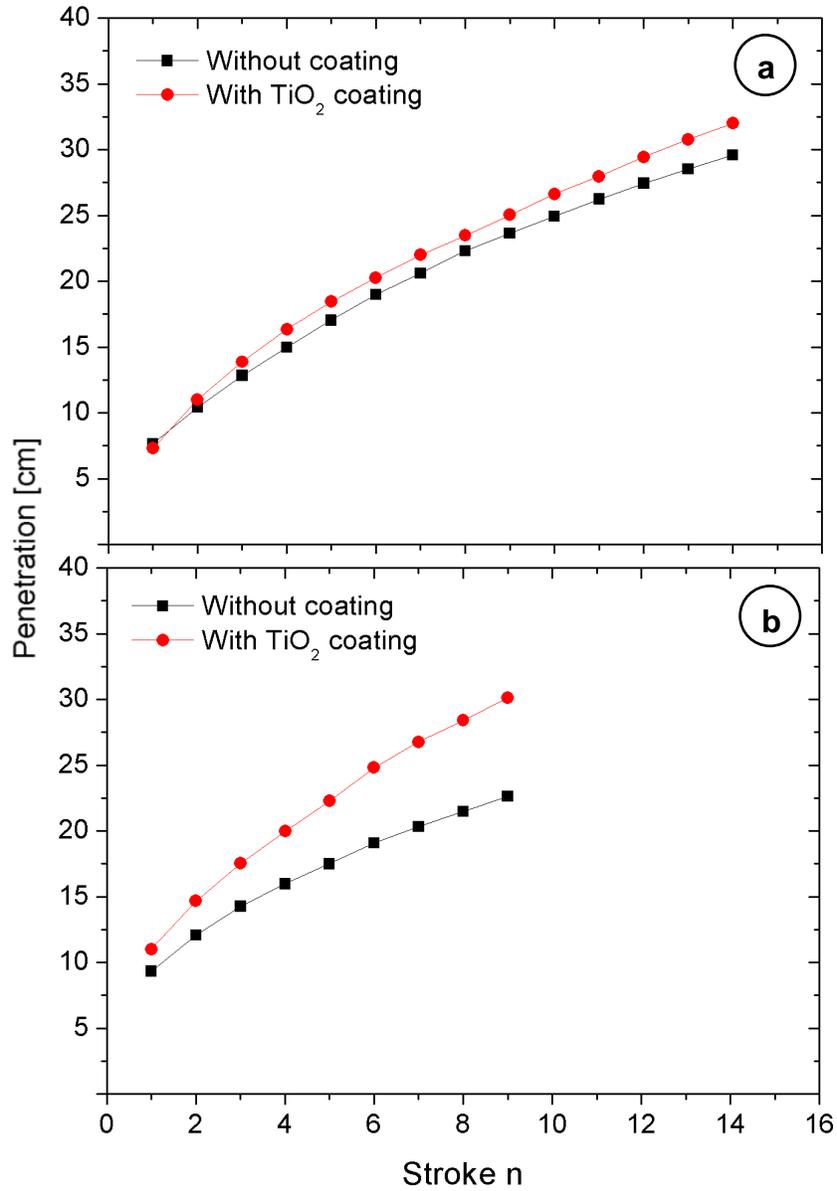

Figure 7. Penetration depths of uncoated and TiO$_2$-coated piles in (a) Plered and (b) Sukabumi clays.

We need to confirm that the difference in penetration depth between uncoated and TiO$_2$-coated piles occurs only in clays and not in regular soils. We performed similar tests on regular soil, regular soil wet 1, and regular soil wet 2. The results of pile driving tests (Figure 8) clearly show that the piles penetrate more readily soils of higher water content. To penetrate regular soil with a pile to a depth of 30 cm requires nearly 60 strokes, whereas for regular soils wet 1 and wet 2 requires 13 and 5 strokes respectively. Similar results were observed for Plered and Sukabumi clay with the pile penetrating the later easily as the water content is higher.

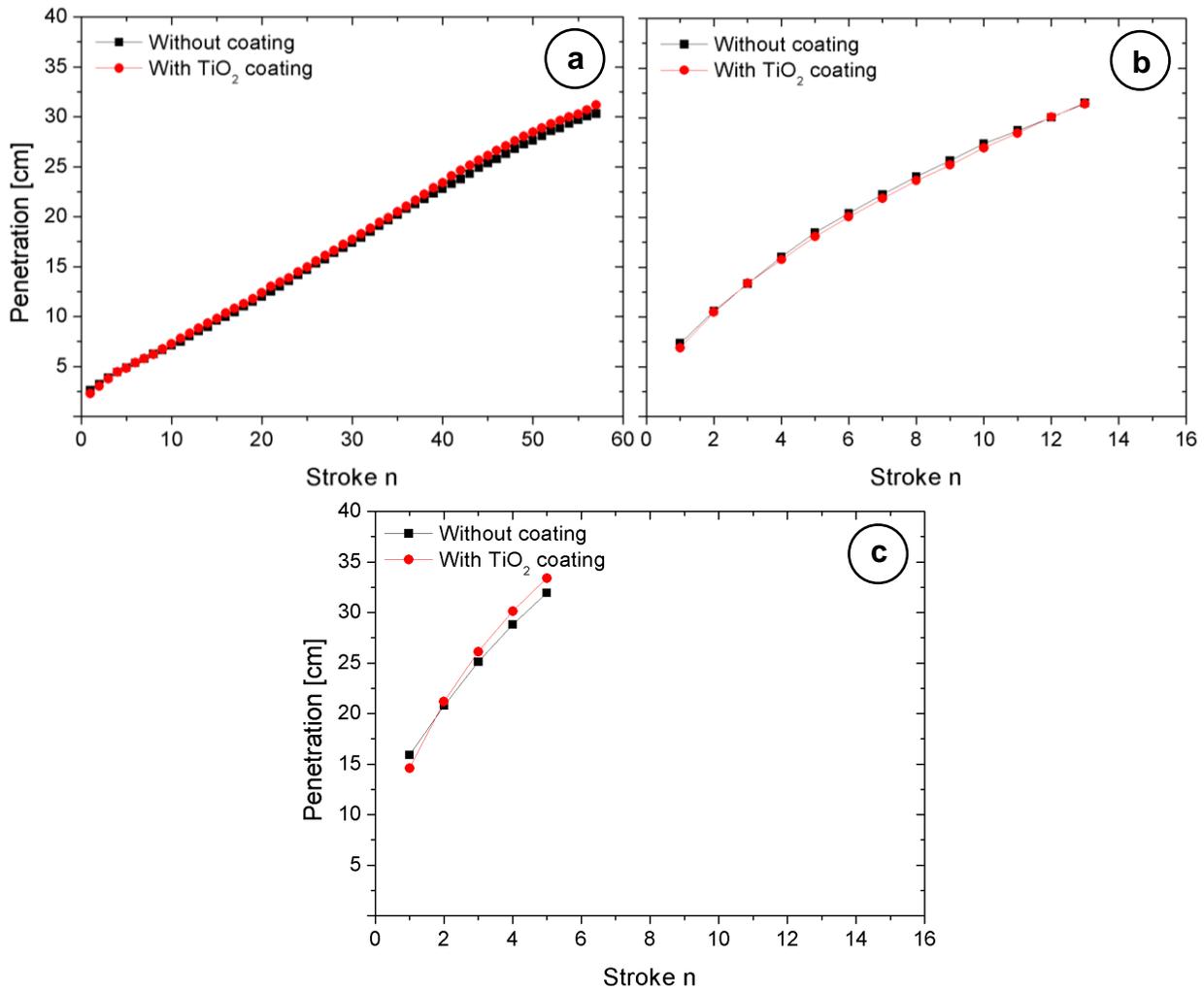

Figure 8. Penetration depths of uncoated and $TiO_2$-coated piles in (a) regular soil, (b) regular soil wet 1, and (c) regular soil wet 2.

The interesting result from Fig. 8 is there is nearly no difference in penetration depths between uncoated and $TiO_2$-coated piles. This indicates an absence of a coating effect in the regular soils. This phenomenon results from weak bonding of water molecules to the particles of regular soil; the water molecules though can move freely between soil particles. If water molecules are attracted to $TiO_2$ particles on the coated piles, the molecules do not affect the movement of the pile as they can move freely between soil particles.

Different situation happened in the clays. The water molecules are attracted strongly by clay particles. Coated piles will be wetted by water whereas uncoated piles will not. Driving these different piles into clay produces different reactions from the clays and hence different penetration depths.

We recorded the FTIR spectra of clay only, concrete only, and mixture of clay and concrete to investigate whether new bonds were being created at the concrete and clay interface. Concrete does not show any FTIR peaks (Figure 9), whereas the spectra of clay and mixture of clay and concrete are nearly similar. The clay spectrum shows stretching and bending of both Si-O and OH in the range of 1300 - 1400 $cm^{-1}$ [13].

The peak location and relative intensity between peaks appearing in the spectra of clay and clayconcrete mixture are identical, indicating that their absorption spectra are clearly identical. This indicates the absence of any new bonds created at the interface of concrete and clay. This result demonstrates likely hydrophobicity or non-hydrophilicity properties because water molecules are not attracted by any atom on the concrete surface.

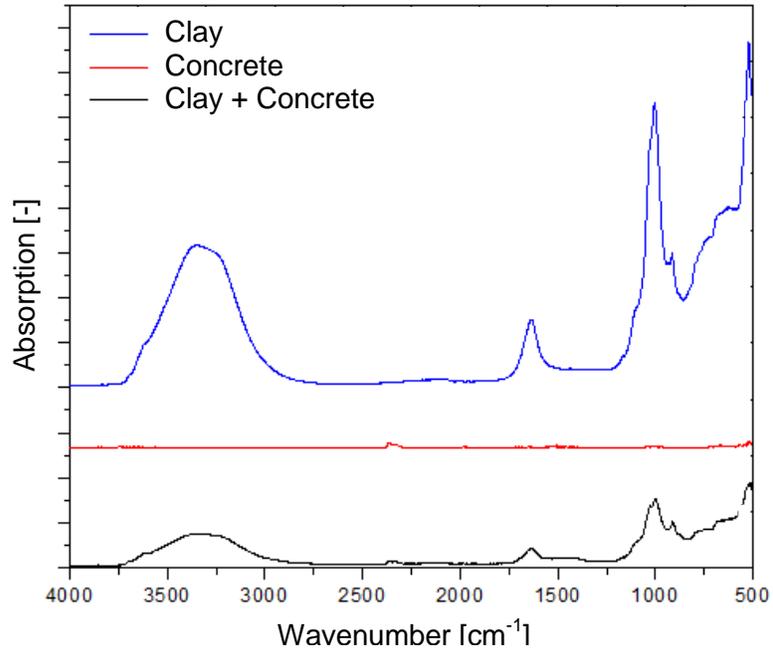

Figure 9. FTIR spectra of clay only, concrete only, and a mixture of clay and concrete.

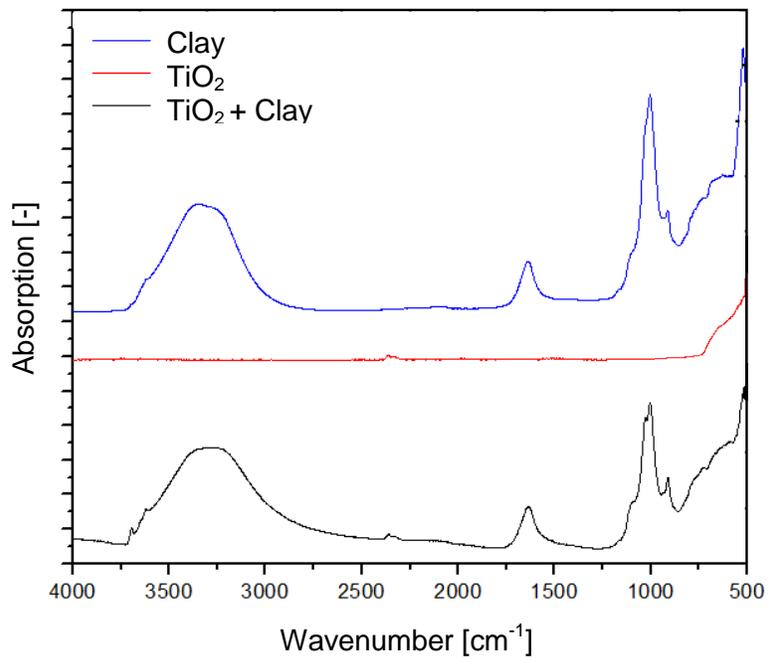

Figure 10. FTIR spectra of clay only, concrete only, and a mixture of clay and TiO$_2$.

Figure 10 shows the FTIR spectra of clay only, $TiO_2$ only, and a mixture of clay and $TiO_2$. The spectrum for the $TiO_2$ powder used is the same as that for standard $TiO_2$ powder with a broad main Ti–O–Ti peak from 500900 $cm^{-1}$ [14]. Comparing the spectra for clay and the mixture of clay and $TiO_2$, no new peaks appear although an interesting phenomenon is observed. The OH band in the spectrum of the clayTiO$_2$ mixture is wider than that belonging to only clay. In the spectrum for the clayTiO$_2$ mixture, the OH band widens up to 2500 $cm^{-1}$, whereas in the clay sample, the OH band spreads up to only 2850 $cm^{-1}$. This widening, attributed to increments in the density of OH bonds in the clayTiO$_2$ mixture, can be understand if in this mixture some water molecules are bound to the surface of $TiO_2$ thereby enhancing its superhydrophilicity. This implies the surface of the $TiO_2$-coated pile is superhydrophilic. The surface of the pile will attract water molecules, halting the compressing of water in the clay pores and ultimately suppressing pile rebound (upward force).

Water has three vibrational modes associated with bond stretching OH ($\approx$ 38003600 $cm^{-1}$ in the liquid state) and bending ($\approx$ 1650–1590 $cm^{-1}$ in the liquid state). The 1638-$cm^{-1}$ band indicates the presence of the OH groups. The position of the molecular band is very sensitive to interactions like hydrogen bonding, which encourages the shift towards lower wave numbers (<3600 $cm^{-1}$), creating differences between water-free, hydrogen-bonded, and intramolecular hydrogen bonds [15].

In Sukabumi clay, the intensive band at 1028 $cm^{-1}$ is related to Si–O stretching. The OH-bending kaolinite demonstrates the presence of surface groups over the OH inner surface at bands 911.6 $cm^{-1}$ and 934.2 $cm^{-1}$ [18]. A vibration band at 911.6 $cm^{-1}$ indicates the possible presence of hematite [19]. The presence of bands at 3690 $cm^{-1}$, 3618 $cm^{-1}$, 2358 $cm^{-1}$, 1636

cm$^{-1}$, 1028 cm$^{-1}$, 911.6 cm$^{-1}$ and 794.1 cm$^{-1}$ indicate the possible presence of illite [20], and bands 3618 cm$^{-1}$, 1636 cm$^{-1}$, 1028 cm$^{-1}$ indicates the presence of gypsum [19].

Two IR bands were also observed in Sukabumi clay at 1645 cm$^{-1}$ and 1628 cm$^{-1}$. These two bands are associated the water H-O-H bending and show that there is water in the soil structure. They are related to hydroxyl stretching wavenumbers, the higher/lower band being related to strong/weak hydrogen bond of water, and both in accordance with the positions of the liquid bending modes [16].

In Plered clay, the band 520.9 cm$^{-1}$ is related to SiOAl (octahedral Al) vibration [13]. The OH-bending kaolinite demonstrated the presence of surface of groups of OH inner surface at 913 cm$^{-1}$ and group of OH surface near 935.6 cm$^{-1}$ [16]. The vibration band at 913 cm$^{-1}$ shows possible hematite presence [17].

Plered clay has weak absorbance peaks at 3696 and 3619 cm$^{-1}$ related to out-of-plane stretching, specifically an external vibration of an OH group and stretching vibrations of internal OH groups [13,16]. These two absorption peaks intensify if the clays interact with concrete, and intensify further if they interact with TiO$_2$. The same pattern occurs in Sukabumi clays with a strong new absorbance peak at 3648 cm$^{-1}$ observed in a mixture of TiO$_2$–Sukabumi clay, which is related to the vibration of external O–H groups.

Finally, we discuss the average friction involved with each strike of the hammer on the pile. The hammer was positioned above the top end of the pile. By assuming linear momentum is constant during the collision between hammer and pile, we have $m_h v_h = (m_h + m_p)v$, with $m_h$ and $m_p$ the masses of the hammer and pile, and $v_h$ and $v$ the speeds of the hammer just before striking the pile and of the hammer and pile just after striking. The kinetic energy of the hammer and pile just after striking becomes $(1/2)(m_h+m_p)v^2 = (1/2)m_h^2 v_h^2/(m_h+m_p)$. The speed of the hammer just before striking the pile satisfies $v_h = (2gt)^{1/2}$, with $t$ is the height through which the hammer falls. The pile and hammer stop after penetrating a distance $\Delta t$. Therefore, the average friction experienced by the pile at each stroke is given by

$$F_{av,n} = \frac{m_h^2 gt}{(m_h + m_p)\Delta t_n}$$

where the index $n$ indicates the $n$-th stroke.

We recorded the displacement of the pile at each stroke and plotted the average friction on piles of different surfaces, i.e., uncoated, araldite-coated, and $TiO_2$-coated, at each stroke (Figure 11).

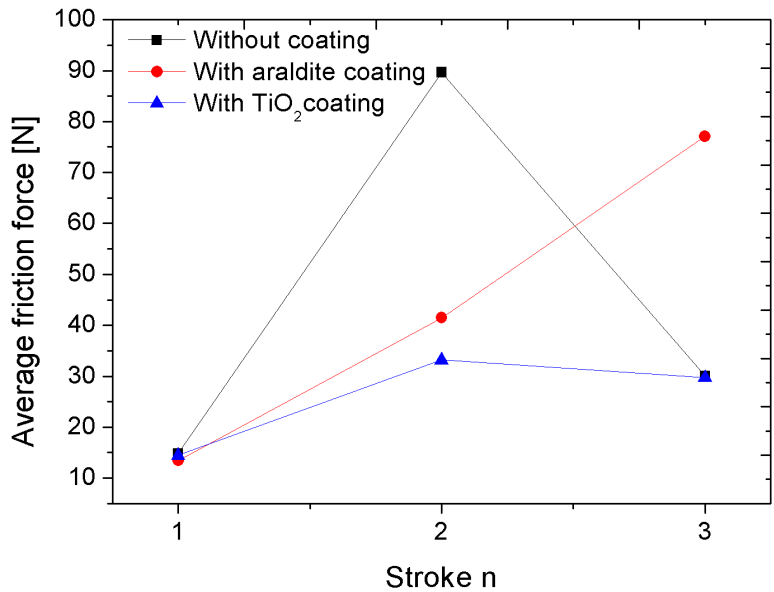

Figure 11. Average friction after each strike of the piles with different surfaces: uncoated, araldite-coated, and $TiO_2$-coating.

Plotting the average friction on piles of difference diameters (Figure 12) shows that the $TiO_2$ coating increases the average depth of penetration. The percentage obtained for each diameter is 1% for the 1-inch pipe, 6% for the 1.25-inch, 14% for the 1.5-inch, and 6% for the 2-inch. Although the percentages range between 1–14%, the difference in penetration depth increases with increasing pile diameter, that is, 2 times for the 1-inch pipe, 3 times the 1.24-inch, 9 times for the 1.5-inch, and 10 times for the 2-inch.

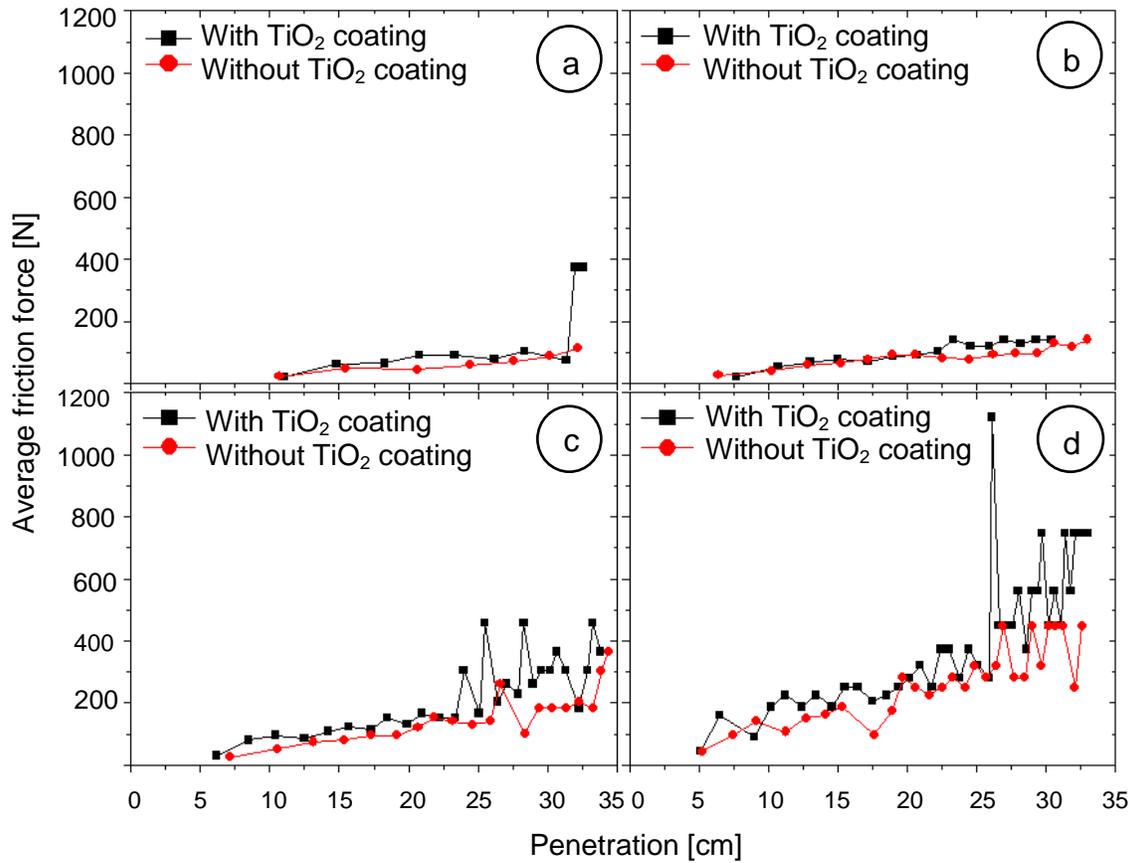

Figure 12. Average friction on piles of difference diameters: (a) 1 inch, (b) 1.25 inches, (c) 1.5 inches, and (d) 2 inches.

The value of the average friction was also estimated analytically yielding increasing values with increasing pile diameter. The analytical model and experimental data show that rebound occurs once at depths of 2325 cm for each pile. The results also indicate that the position of a high pile rebound is independent of the size of the pile diameter. The effect of rebound has been reduced by up to 50% using $TiO_2$-coated piles.

**Conclusion**

We have demonstrated that changing the behavior of concrete pile surfaces to very hydrophilic or superhydrophilic by coating with $TiO_2$ nanoparticles enables easy penetration of such pipes in clays and reduce pile bouncing. Compared with uncoated piles, the $TiO_2$-coated piles for any size penetrated deeper into the two clays used in this work: around 7.3% in Plered clay and 22% in Sukabumi clay. However, in regular soil and regular soils with higher water content, piles with a surface coating showed nearly no effect on pile penetration depth. Interactions between water molecules in the clay pores are believed to be the origin of pile bouncing. Bouncing occurs when the pile surface is hydrophobic and absent if the pile surface is nearly superhydrophilic.

**Acknowledgements**

This work was supported by PMDSU research Grant No. 314d/I1.C01/PL/2015 and BPPDN Research Grant No. 310y/I1.C01/PL/2015 from the Ministry of Research and Higher Education, Republic of Indonesia, 2015.

2317) Gadsden, J.A., Infrared Spectra of Minerals and Related Inorganic Compounds, London: Butterworth & Co, 1975.

18) Wolf, R.G., "Structural effects of kaolinite using infrared absorption," American Mineralogist, Vol. 48, 390, 1963.
23